# What are AI researchers worried about?

Cian O'Donovan,* Sarp Gurakan, Ananya Karanam, Xiaomeng Wu, Jack Stilgoe

Department of Science and Technology Studies
University College London, United Kingdom

March 2026

**ABSTRACT**

As AI attracts vast investment and attention, there are competing concerns about the technology's opportunities and uncertainties that blend technical and social questions. The public debate, dominated by a few powerful voices, tends to highlight extreme promises and threats. We wanted to know whether public discussions or technology companies' priorities were representative of AI researchers' opinions. Our survey of more than 4,000 AI researchers is, we think, the largest conducted. It was designed to understand attitudes to a variety of issues and include some comparisons with public attitudes derived from existing surveys. Most previous surveys of AI researchers have asked them for predictions (Ahmed et al 2024), for example, of passing a technological threshold or the probabilities of some catastrophic event. These surveys mask the uncertainty and diversity that normally characterises scientific research.

Our hypothesis was that the opinions of AI researchers would be markedly different from those of members of the public. While there are areas of divergence, particularly in attitudes to the technology's potential benefits, our survey shows some surprising convergence between researchers' and publics' opinions, particularly in the assessment and prioritisation of risk. Responses to an open text question 'What one thing most worries you about AI?' reveal that only 3% of AI researchers prioritise existential risks, despite the prominence given to these risks in media and policy. AI technologies and AI researchers seem to be much more 'normal', in the sense used by Narayanan and Kapoor (2025), than public representations suggest. Current discourse about AI may be contributing to public alienation and creating a barrier to public participation. Our survey results suggest the possibility for new forms of public dialogue on AI's risks and opportunities. Rather than speculating on future potential risks, policymakers and AI researchers should collaborate on understanding and mitigating the range of sociotechnical risks that are already of clear public concern. Our survey took place at a time when AI research was becoming rapidly privatised, raising the risk that the space for constructive dialogue between researchers and the public may quickly close.

## 1 INTRODUCTION

Advances in artificial intelligence research [17, 31] have contributed to an explosion of interest and investment in AI technologies. The benefits and the risks of the technologies remain uncertain and contested, but these technologies are already immensely powerful. This raises profound governance questions. The most advanced 'frontier' models are very large, concentrating power in the hands of a few giant technology companies. This power is not just exercised financially. There is also a concentration of discursive power [27] - the ability to shape debates. The voices of a few technology leaders and prominent scientists are amplified, and the terms of debate are simplified. We hear relatively little from the scientists who are working directly on developing and seeking to understand these new technological possibilities.

With other technologies, it has been common for scientists and policymakers to put public mistrust down to ignorance or misunderstanding of science, although we know from social research that this explanation

---

* Corresponding author. cian.o@ucl.ac.uk UCL Centre for Responsible Innovation, Department of Science and Technology Studies, University College London, Gower Street, WC1E 6BT, United Kingdom

is flawed [35]. With new technologies, relevant evidence for risk assessment is typically uncertain. Evidence from the sociology of technology predicts that those closest to a technology are likely to be more uncertain about that technology than those at one stage removed [18]. We should therefore be sceptical of those assessing technologies if they have political or financial stakes. With AI, we have seen some surprising assessments of risk. The emerging science of 'AI safety' has recognised a variety of risks [11], but has prioritised existential, long-term and speculative risks over those that are more immediate and unequally distributed [2]. We would predict from earlier studies in feminist sociology that marginalised groups of researchers are likely to prioritise issues and research agendas differently, in part due to their standpoints [12].

Here, we define risks broadly, following Rosa, as possible threats to things that people value [26]. Alongside risks to health and the environment, we are also interested in concerns about potential risks to livelihoods, lifestyles, markets, democracies and other social institutions (what Jasanoff calls social and political risks [15]). For example, in our survey, we asked researchers whether AI companies should be allowed to train their models on any text or images that are publicly available, in response to real-time policy concerns about potential copyright infringement [16, 32]. With new technologies, issues and controversies can play out very differently in different societies and jurisdictions [19, 20, 30]. Given that important regulatory structures are already being built on risk-based lines [8], it is vital to interrogate the prioritisation of these risks among groups, including researchers (see also Wirz et al. in the case of synthetic biology [34]).

## 1.1 Surveying AI researchers

We wanted to know if and how the views of AI researchers differed from those of members of the public. There have been many surveys of public attitudes to AI, some of which are motivated by a governance model that problematises and tries to anticipate public ignorance and mistrust. There have been relatively few surveys of AI researchers' views. Some of those that have been published have been widely reported, despite having poor survey design or small samples.

A 2023 "Expert Survey on Progress in AI," of 2,778 AI researchers, recruited via contributions to major conferences and journals, was framed in ways that undermine the value of its findings [10]. Respondents were treated as possessing privileged access to relevant knowledge and asked to forecast AI developments and implications. The survey, part of a series run by an organisation with a focus on existential risk, asked respondents to put a probability on AI developments "causing human extinction or similarly permanent and severe disempowerment of the human species?" [10]. This probabilistic, predictive and expert framing sheds very little light on diversity or uncertainty among AI researchers and doesn't allow for alternative framings of risk or concern.[1]

Some surveys, rather than treating researchers as experts with privileged access to truths that might enable prediction, have instead taken a reflexive approach, asking scientists to consider their own field. A smaller survey of 475 members of the Association for the Advancement of Artificial Intelligence sought clarity from AI researchers on issues that they felt were being poorly represented in public. It found, for example, a majority of respondents (76%) agree that "scaling up current AI approaches" is "unlikely" or "very unlikely" to produce artificial general intelligence (AGI) and "The majority (77%) of respondents prioritize designing AI systems with an acceptable risk-benefit profile over the direct pursuit of AGI (23%)".

Bao et al. have conducted a more robust survey of AI scientists (n= 2,352), using a latent profile analysis to identify patterns of risk and benefit perception [4]. They further ask whether these perceptions are linked to measures of political ideology, degree of deference to scientific authority and news consumption. They also ran comparisons with a public survey. They found that scientists were substantially less likely than members of the public to be indifferent (0.2% versus 3.1% of the public), sceptical (9.5% versus 25.3%) or ambiguous (28.8% versus 45.9%) and much more likely to be positive about AI (20.6% versus 4.4%). Interestingly, they also found scientists were much more likely to be ambivalent (40.9% of scientists versus 21.2% of the public), which they define as perceiving high risks and high benefits from AI. They conclude that there needs to be more fine-grained analysis of attitudes to potential risks and benefits, including open-ended survey questions.

---

[1] For an interesting response on the methodological choices taken by this survey and our one, see [9].

A further survey from the Pew Research Center compared public attitudes with the views of 1,013 respondents referred to as 'AI experts' [25]. This survey asked in general how positive people felt about AI's impacts on various aspects of society, and included questions about governance and responsibility. Among its findings was that, despite experts' increased optimism about the technology, there was a shared desire between experts and the public for increased regulation.

## 2 METHODS

### 2.1 Sample

There was no readily available list of AI researchers that could serve as a sampling frame for our survey. To recruit participants recently active in AI research and related fields we identified candidate participants on the online academic article repository ArXiv. We identified articles published in the following AI-related computer science subcategories: cs.AI (artificial intelligence), cs.LG (machine learning), cs.CV (computer vision and pattern recognition) and cs.CL (computation and language). This includes researchers in different branches of AI, researchers from other science and engineering fields that applied AI tools and methods, and researchers from social science who study AI and its applications. We identified articles submitted between January 1st, 2020 and February 23rd, 2024. We built a corpus of 43,325 articles from which we identified 165,226 unique authors.

We took a purposive, convenience sampling approach to gain a deep understanding of AI researchers views on responsibilities. This is similar to the approach taken by Bao et al. (2025). However, while they constructed a population of article authors using data from Web of Science, we used ArXiV as it is a pre-print platform popular with non-academic researchers, including those in firms and the public sector.

Using the Qualtrics survey platform we sent an initial invitation email and two follow-up reminders to a list of 99,516 authors of articles in our corpus, randomly selected. No incentives were provided to potential respondents. The survey was open for a month, between June 20th and July 22nd 2024. At the end of this period, 7,595 surveys had been started and 5,318 of those completed, a response rate of 7.6% [3 RR6]. This response rate is in line with that of Bao et al. (8%), who used similar purposive sampling techniques to construct a sample of AI scientists [4]. It's interesting to note here that while Bao et al. contacted only corresponding authors, we included all paper authors in our invitee list. This decision seems to have had a negligible impact on response rates. We considered for our analysis only completed surveys and those that had clicked 'yes' on the pre-survey consent form. The final dataset for analysis contained 4,260 responses. The study design was reviewed and awarded research ethics clearance by the UCL Department of Science and Technology Studies Research Review Panel.

Responses to demographic questions reveal that in our sample, 77% of our respondents work in university or research organisations, while 23% are in industry or 'government / public sector'. 71% of respondents have PhDs, and another 22% have some graduate degree. 75% replied 'mostly' to the question 'to what extent does your job involve AI?' A further 20% replied 'somewhat', 4.7% replying 'slightly' and 32 responding 'not at all'. 81% of respondents described their gender as male and 18% female. 40 respondents identified as non-binary, 10 self-described their gender and 117 chose not to identify with a given gender or add a response of their own.

Asked to specify their country of work, respondents (n=3,765) listed 92 countries. The top countries are the USA (25%), the UK (11%) and China (8%). 29% of respondents live in European Union countries. 714 respondents, (17%), work in one of 47 low and middle-income countries entered. Of this number 60% listed

China and India. It is important to note that our survey was available only in English, and that the population of researchers we sampled had also published their papers in English.

## 2.2 Survey design

Our overarching research questions were:

RQ1. What do AI researchers think of AI risks, benefits, responsibilities and other issues?
RQ2. Where are the gaps and overlaps between public and AI researchers' attitudes?
RQ3. How do AI researchers imagine public engagement?

Our focus here is on a block of questions about risks, concerns and worries.[2] This included clickable lists of possible risks and benefits taken from a public survey, and a number of agree-disagree Likert scale questions. In addition, we asked an open text question "What one thing most worries you about AI?"

To enable comparison, some of our questions were taken from existing surveys of UK public opinion, by the Ada Lovelace and Alan Turing institutes [29] and by the Office for National Statistics [23], both in 2023. Our survey of researchers was global but these public surveys were national. Comparisons therefore should not be read too precisely. We ran checks to see if UK-based AI researchers differed markedly on any major issues and found no substantial divergences.

## 2.3 Coding free text responses

The 3,718 non-blank responses to the open-ended survey question "What one thing most worries you about AI?" (Q2.2) were coded manually (after an initial experiment with using BERTtopic modelling software proved unhelpful). One researcher's initial categorisation of 10% of the responses produced 65 codes (categories), situated under 35 higher-level codes. After discussion and refinement with the research team, the total set of valid responses were coded, producing 51 codes (including N/A and uncategorized), with some responses assigned multiple codes, up to a maximum of three. The coding was informed by expert researchers with an understanding of discourses among AI research that were in circulation in the summer of 2024. So, for example, responses related to the "black box" nature of AI models were categorised alongside those concerned about explainability, interpretability, and transparency.

Quantitative analysis was performed by translating codes into counts and prevalence. To examine correlations between code prevalence and other survey variables, a two-sample proportional z-test was used with a statistical significance threshold of $p < 0.05$. This test was selected because it is well-suited for comparing independent proportions across two groups [28:229], which fits the structure of the data, for instance, comparing the proportion of male versus female respondents expressing worries using a specific code. The analysis was run selectively on five sets of theoretically relevant variables drawn from key survey questions: gender, sector, inevitability of artificial general intelligence, deficit model of the public, and optimism. Each captures a demographic group or dimension of interest expected to shape how respondents articulate concerns about AI. For instance, sector (respondents working in universities, firms or the public sector) was chosen because we hypothesized that differences between organizations would influence how researchers interpret worries around AI, given divergent institutional contexts and incentives.

## 3 SURVEY RESULTS

### 3.1 Risks versus benefits

When asked about the balance of AI's risks and benefits, researchers are positive. A large majority of researchers (87%) believe the benefits either outweigh or are balanced by the risks. Only 9% believe the risks outweigh the benefits. In comparison, the response profile from an ONS study that surveyed members of the UK public shows a slight majority of the public (57%) believe the benefits either outweigh or balance risks. 28% of the UK public believe the risks outweigh the benefits.[3] When asked about particular benefits, researchers are enthusiastic about potential improvements in education and work (Figure 1).

---

[2] The full set of survey questions and commentary on responses is available at [22].

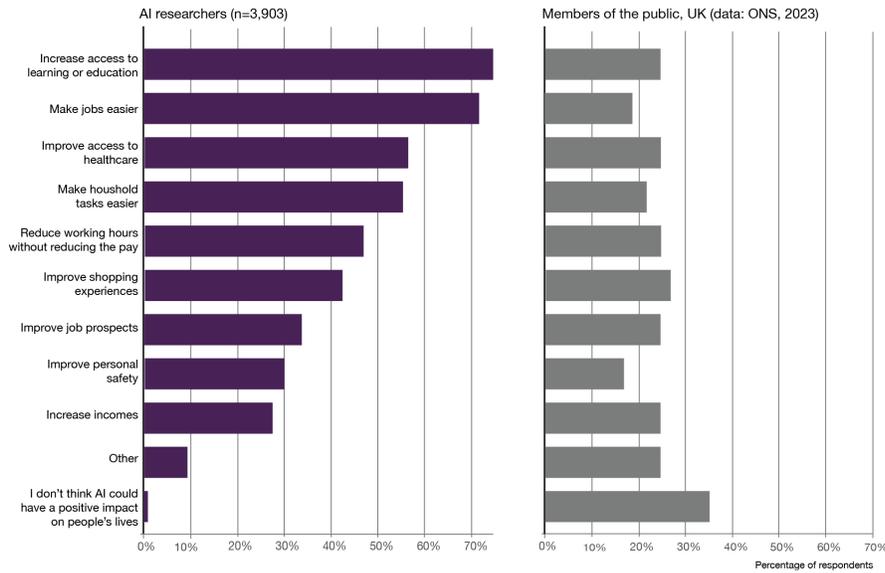

Figure 1. 'In which ways, if any, do you think artificial intelligence could have a positive impact on people's lives?'

There is substantial divergence between AI researchers and the public in their perceptions of AI's benefits.[4] Less than 1% of AI researchers agreed with the statement 'I don't think AI could have a positive impact on people's lives'. In comparison, 36% of respondents to the 2023 ONS survey of members of the public in the UK do not believe AI will have a positive impact on people's lives.

## 3.2 Prioritising harms

There was notably more agreement between AI researchers and the public when it came to the perceived harms of AI (Figure 2).

---

[3] Previous surveys of emerging technologies find that publics see higher risks and lower benefits than scientists for these technologies [13, 14].
[4] For this question we again compared AI researcher views from our own survey with responses to the 2023 ONS survey of members of the public in the UK [23].

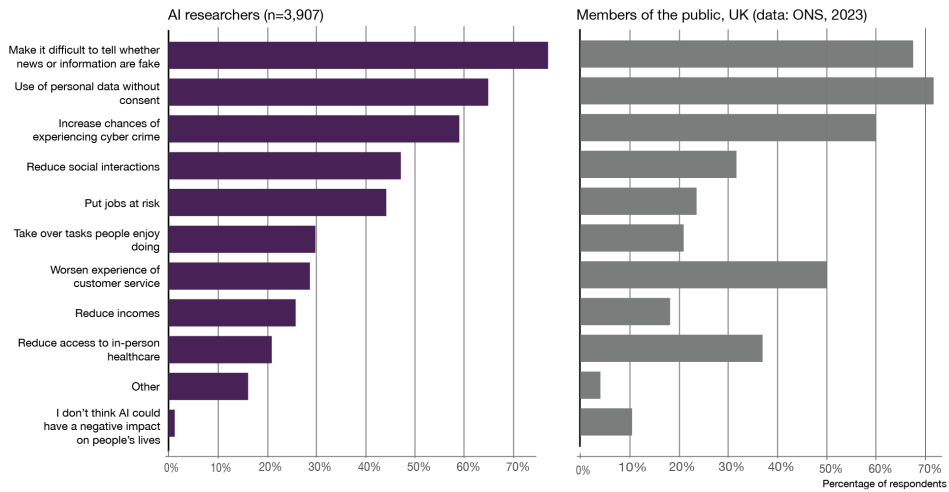

Figure 2. In which ways, if any, do you think artificial intelligence could have a negative impact on people's lives?'

When asked to select negative impacts AI could have on their people's lives, researchers and the public agreed on the top three: disinformation, use of data without consent, and cybercrime. Perspectives diverge after that.[5] For the public, the fourth agreed concern is worsening experiences of customer service. For AI researchers, it is reduced social interactions. What is notable here however is the relative consensus between researchers and members of the UK public when it comes to negative impacts of AI, compared with an 'optimism gap' when it comes to benefits.

A large majority of researchers were concerned that both risks and benefits would be unequally distributed (Figure 3).

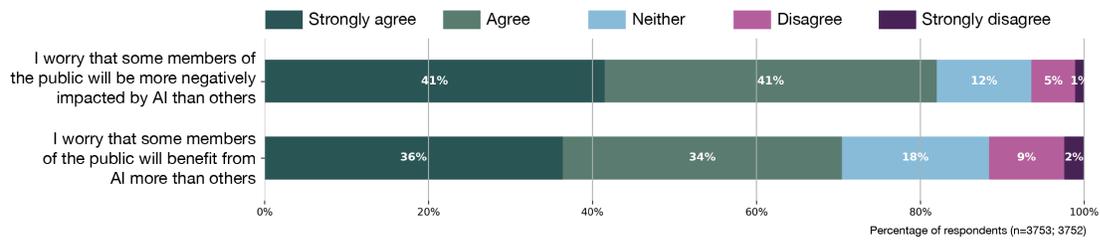

Figure 3. 'To what extent do you agree with the following statements? - I worry that some members of the public will be more negatively impacted by AI than others; I worry that some members of the public will benefit from AI more than others'

Almost all agreed that regulators should consider impacts on different groups, but this was significantly more pronounced for female researchers (Figure 4).

---

[5] For this question we again compared AI researcher views from our own survey with responses to the 2023 ONS survey of members of the public in the UK [23].

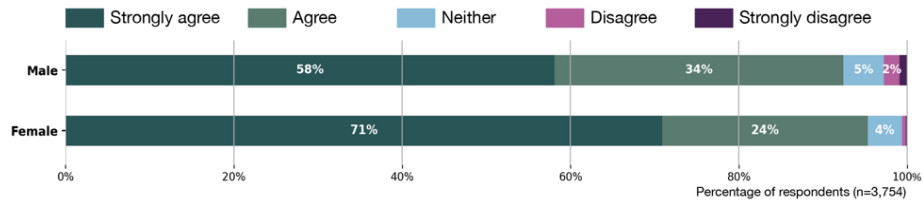

Figure 4. 'To what extent do you agree it is important that regulators consider how AI impacts different groups?'

## 3.3 Free-text concerns

As well as asking AI researchers to choose among pre-selected risks and benefits, our survey also sought bottom-up responses, asking AI researchers 'What one thing most worries you about AI?' The free-text responses to this question were coded and clustered to generate a typology of concerns.

These concerns are listed in Table 1. This is a broad range of concerns, with no single issue dominating. Notably, only 3% of researchers stated they were most worried about existential risk–offering a corrective to the "AI safety" focus on long-term, speculative risks. This small proportion is in contrast to a diverse array of near- and medium-term societal impacts, some of the most prominent being malicious use (11%), misuse (10%), misinformation (9%) and jobs (7%). This indicates that researchers tend to view AI systems as "normal technology" rather than potential superintelligence, worrying primarily about their deployment and resulting harms [21].

Table 1. Categories constructed from responses to the question: 'What one thing most worries you about AI?'

| Topic | Freq. | Description | Example response text (verbatim) |
|---|---|---|---|
| malicious use | 10.6% | Intentional use for harm by members of the public or criminals | "How people (in particular bad actors, armed forces and profit-hungry corporations) make use of AI. Let me rephrase that: I am not worried about super intelligent gone-rogue AI itself (like in a HAL9000-like or SkyNet-like scenario), but about how people deploy way more "stupid" AI for their own benefit..." |
| misuse | 9.9% | Incorrect or inappropriate use of AI | "Its a powerful tool which can be misused in the wrong hands" |
| misinformation | 8.8% | AI spreading false or misleading content | "Fake news and videos being created for social media and influencing people e.g. in how to vote. " |
| impact on jobs | 7.1% | AI replacing human labour or reducing employment opportunities | "Massive unemployment, as this time there may be no way out: this is the first time that AI may replace white collar jobs entirely" |
| public understanding of AI | 5.4% | Non-experts misunderstanding AI systems | "The inability of people to recognize that current AI is not intelligent but a fancy description of various statistical methods" |
| general societal impacts | 5.2% | Reshaping of relationships, norms, and social structures due to AI | "Its social impact - further polarisation/ division within societies." |
| hype | 5.1% | Overpromising or exaggerated claims about AI | "People overhype it or else see it as an evil in and of itself, as if we lived in a sci-fi movie. Rather than being generally informed about how it works. Just like most people should know generally how a washing machine works or their car." |
| bias | 4.5% | Algorithmic bias and unfair outcomes from AI systems | "How pattern detection, particularly in biased, historical data, will lead to propagated bias, discrimination, and deepening inequality within and across societies in the world. " |
| privacy | 4.0% | AI compromising or exploiting private information | "There can be privacy concerns regarding the sensitive information of users' data used to train the AI" |
| performance | 4.0% | Hallucinations and other limits in accuracy and robustness | "The fact that AI can very confidently convey wrong information." |
| regulation | 4.0% | Too little, too much, or problematic AI governance | "poorly thought out policy making resulting from hype and misconceptions" |
| alignment and control | 4.0% | Ensuring AI aligns with human values/control | "Making increasingly high-stakes decisions using AI without knowing that is aligned with our interests" |
| power concentration | 3.9% | Control of AI by a few dominant actors | "Centralizes power in the hands of a few corporate actors, through its (material) dependencies - compute, data, labor, software frameworks. " |

| Topic | Freq. | Description | Example response text (verbatim) |
| --- | --- | --- | --- |
| inequality | 3.8% | Unequal distribution of AI benefits/harms | "I think that it has the ability to make billions for 0.001% of the population and destroy the lives of millions of others without compensation for the losses" |
| black box | 3.8% | Lack of transparency or explainability in model reasoning | "That they are almost always blackboxes. We still lack a strong theoretical framework for what happens inside an AI." |
| existential risk | 3.4% | Fear that advanced AI could cause catastrophic or irreversible harm | "Rogue AI taking over, human extinction" |
| overuse | 3.3% | Overreliance or unnecessary use of AI | "That seemingly everybody wants to use it for everything, even where AI does not make sense at all" |
| profit-driven | 3.2% | Development of AI for private, commercial priorities over public interest | "It's owned and governed by people who are only interested in money and power; it's usually not put to use for the public good; more intelligent will likely mean more risk; focuses on the wrong goals (it helps me with my intellectual tasks, while I want AI to do my laundry, cleaning, etc - or otherwise solve world problems like war and climate)" |
| irresponsibility | 3.0% | Absences or shifts of responsibility by developers/users | "Irresponsible deployment of AI to the real-world." |
| uncritical use | 3.0% | Blind trust or careless use of AI systems | "Blind use of AI, without understanding of its limitations, especially for deciding on social issues (i.e., AI for deciding on policy, predicting crime, deciding what's biased or toxic, etc.)" |
| ethics | 2.4% | General ethical or moral concerns about AI | "AI does not think and has no ethics or morals." |
| deskilling | 2.3% | Loss of skills and critical thinking due to AI reliance | "I may become lazy and lose some parts of my skills because I ask AI to do it." |
| tech companies | 2.1% | Critiques of Big Tech control and behaviour | "The strongest AI teams are usually part of big tech companies whose objectives and interest do not align with the well-being of all humans but with their own economic benefit." |
| human in the loop | 1.9% | Lack of human oversight in AI decisions | "Widespread delegation of important decision-making to AI algorithms with little human involvement." |
| manipulation | 1.9% | AI used to influence or control people | "Large-scale political manipulation" |
| advanced AI | 2.0% | Development of superintelligent or sentient systems | "I think the most interesting one is AGI." |
| general concerns | 1.7% | Broad or unspecified AI worries | "Unexpected negative outcomes" |
| n/a | 1.5% | Responses that do not meaningfully address the question | "No." |
| safety | 1.6% | Unspecified fears about AI being unsafe | "AGI is approaching but we are not ready in terms of safety." |
| expert understanding | 1.6% | Whether experts truly understand AI systems | "People (even researchers and practitioners) inability of understanding its limits, risks and benefits and its potential social impact, both positive or negative." |
| speed | 1.6% | Rapid development of increasingly large models | "The speed at which the industry is growing and adapting AI in almost everything. It is really hard now days to separate facts from fiction." |
| energy and resource | 1.4% | High electricity/water consumption requirements | "Obscene energy use, increasing technocapitalist domination of life" |
| nothing | 1.4% | Respondent explicitly expresses no concern about AI | "Nothing, it's just a software." |
| environment | 1.3% | Environmental harms from AI development | "It's carbon footprint and the potential for us to exacerbate global warming to the point society fails" |
| security | 1.2% | AI creating or exposing security risks | "Security vulnerabilities can be easily exploited and people rarely seem to care" |
| slop | 1.1% | Low-quality mass AI content polluting online spaces | "A deluge of generated disinformation, misinformation, and low-quality content displacing quality works by humans." |
| training data | 1.1% | Issues with data sourcing, labelling, and attribution | "Predatory nature of obtaining data for model training abusing rights of millions of creators" |
| slowing progress | 1.0% | Concerns that AI progress may stall or pause | "The development of AI systems may get stalled after some years." |
| creativity | 1.0% | Loss or devaluing of human creativity due to automation | "Coming generations will lack creativity" |

| Topic | Freq. | Description | Example response text (verbatim) |
| --- | --- | --- | --- |
| disempowerment | 1.0% | Loss of human agency or meaning due to automation | "Eventual disempowerment of humans as the agency, capability, and generality of AI systems increases" |
| fraud | 0.9% | AI used for scams or deception | "Fraud using AI, such as deepfake and voice imitation" |
| AI research ecosystem | 0.9% | Concerns about AI research norms/direction | "Leading edge of research is not in universities" |
| specific applications | 0.7% | Risks tied to particular domains (health, law, etc.) | "Application of LLMs to the medical domain" |
| transparency | 0.6% | Lack of clarity about model development/use | "Lack of transparency in the development practices and the kind of data used to train widely used models" |
| indistinguishability | 0.7% | Hard to tell AI versus human content | "Not being able to distinguish whether I'm interacting with humans or machines…" |
| costs | 0.6% | High financial and computational costs | "The computation of large models. Only the rich and big organizations can afford the GPUs and the costs, leaving the rest of people from research and development…" |
| anthropomorphisation | 0.5% | Attributing human traits to AI | "The Eliza effect: attributing too much ability and responsibility to AI" |
| data contamination | 0.3% | AI-generated data polluting training datasets | "Degrading quality due to training data being itself AI generated" |
| geopolitics | 0.2% | AI driving international competition/conflict | "That it leads to a new cold war-era where geopolitical tensions are high and conflicts are numerous." |
| participation | 0.1% | Lack of public involvement in AI governance | "AI systems are deployed without asking opinion of general public and impact of such systems" |

There is a notable cluster of concern around downstream uses–what happens when AI models leave research labs and end up in the hands of users. Nearly 25% of responses referenced one or more of "malicious use", "misuse", "overuse", and "uncritical use." The frame here is a familiar one in which the technology and its creators are regarded neutrally while the uses of technology are problematised. These can be contrasted with smaller but still significant numbers of what we might call reflexive concerns - about the technology itself or its innovators. These concerns include hype, the performance of AI, concentration of power, the provenance of training data, 'black box' inscrutability and profit motives.

Some of these concerns are strongly correlated with other responses to demographic or attitudinal questions. Women are much more likely than men to worry about bias. 8% of women expressed this concern rather than 4% of men. Researchers who think AGI is inevitable are much more likely (5% versus 1% of AGI non-believers) to worry about existential risk/alignment and much less likely to worry about hype (2% versus 10% of AGI non-believers). Belief in the inevitability of AGI was associated with a significantly higher rate of expressing worries about privacy (5% versus 2%, $p = .00002$). While a z-test only establishes a proportional difference rather than causality, this gap underscores how different imaginaries of AI trajectories shape present-day concerns.

### 3.4 Other concerns

In recent years, we have seen growing attention to issues around AI that aren't just related to direct risks or benefits. In our survey, we asked AI researchers questions about explainability, training data and responsibility for the governance of AI.

*3.4.1 Explainability*

A frequently-highlighted concern about AI is that technologies may be 'black boxes' [24]. Simply put, it may be hard for a combination of technical and social reasons, to fully understand why an AI system does what it does [5]. Public surveys have sought to identify values in a straightforward trade-off between the performance of a systems, expressed in terms of accuracy and an explanation of decision-making [29]. When compared with responses from members of the British public [29], AI researchers are on average less concerned than the public are about the need for explanation (Figure 5).

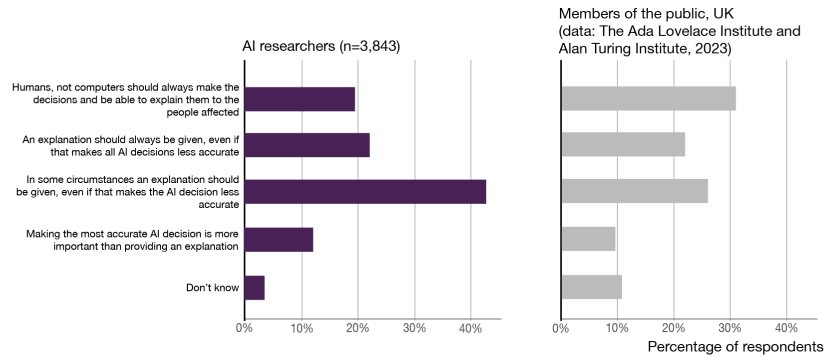

Figure 5. '[Here are] four statements that reflect different opinions toward explaining how AI systems make decisions. Overall, which statement do you feel best reflects your personal opinion?'

*3.4.2 Training data*

AI researchers in industry and in universities were concerned about the data on which AI companies trained their models. Only one in four of the survey's respondents think that AI companies should be allowed to train their models on any publicly available text or images. More than 65% agree that there should be some sort of constraints, either explicit permission, or opt out systems (Figure 6). Responses did not vary significantly across industry and academic sectors. 47% of researchers working in academic roles told us 'AI companies should only be allowed to train their models on text or images where they have explicit permission to do so from the original creator' while the number for researchers working in industry was 43%. While the numbers of members of the public in the UK opting for this kind of explicit consent was lower (37%), this is in part explained because 25% of the public responded with 'don't know' [7] (compared with only 5% of respondents in our survey.

*3.4.3 Responsibility and public participation*

At a time when governments around the world are working out how to regulate emerging AI systems, AI researchers' views reveal a desire to take responsibility. It has been easy in the past for those involved in developing or researching a technology to offload responsibility for the technology's impacts onto its users. We were surprised to see substantial agreement with a strong idea of accountability, two thirds of researchers agreed that the people who create AI systems should be held responsible for their real-world impacts (Figure 6).

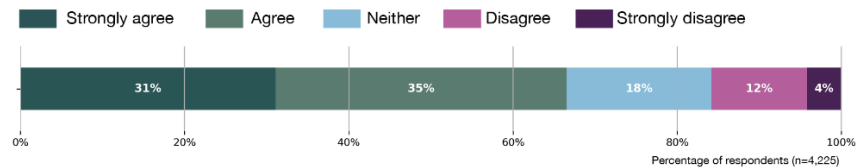

Figure 6. To what extent do you agree or disagree: The people who create AI systems should be held responsible for the real-world impacts of those systems'

*3.4.4 The direction of artificial intelligence innovation*

AI researchers were concerned not just with what AI technologies might do in the world or how there were constituted and governed, but also with questions of the trajectory of AI. A majority of AI researchers were concerned that priorities were currently being controlled by industry and most wanted AI models to be open source (Figure 7). Recognising that 'open source' is a contested phrase in AI, we offered no definition in our survey, so the views here should be seen as hinging on varying perceptions (positive, negative, technical etc.) of the term [33].

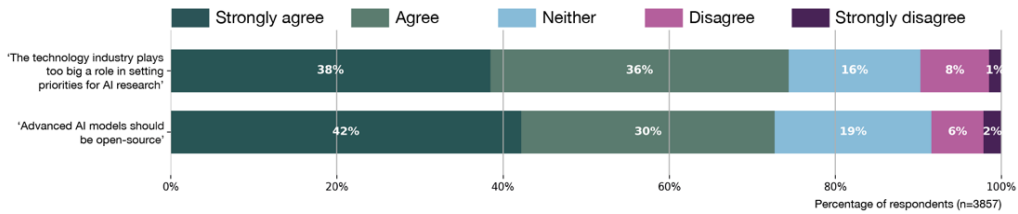

Figure 7. 'To what extent do you agree or disagree with the following statements. The technology industry plays too big a role in setting priorities for AI research; Advanced AI models should be open-source'

When asked about funding for different areas of AI research, researchers express clear preferences for healthcare and education, with much weaker enthusiasm for military AI (Figure 8).

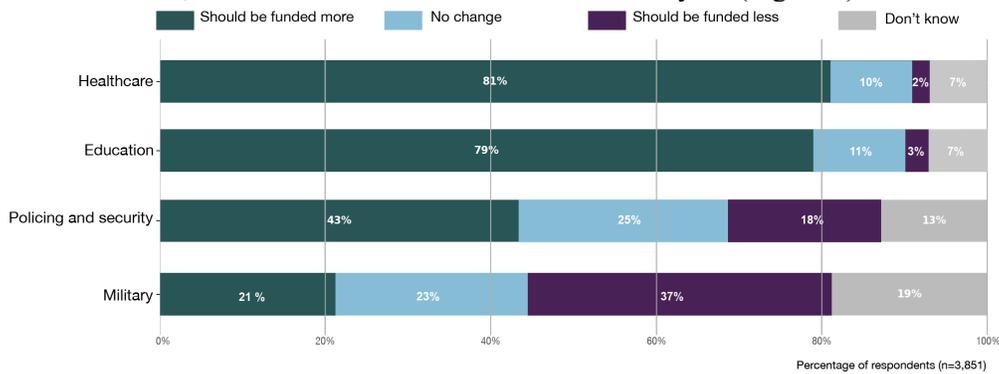

Figure 8. 'How should government funding for AI be changed in these areas?'

Given the prominence of debates about 'artificial general intelligence' in AI research,[6] we wanted to know whether researchers thought AGI was inevitable and how views of inevitability might relate to other views. 51% agreed or strongly agreed that 'artificial general intelligence (AGI) is inevitable', 24% disagreed or disagreed strongly, while 25% responded 'neither'.

To a statistically significant degree, researchers who view AGI as inevitable are much more likely to think that the technology should be developed as quickly as possible (Figure 9). Conversely, researchers who do not think AGI is inevitable are significantly more likely to list hype as their primary worry, and much less likely to mention worries about privacy, alignment or existential risk when compared with other researchers (see Table 1).

---

[6] Definitions of AGI vary, and the term is often-discussed but highly contested within AI research. We offered no definition in our survey, so the views here should be seen as hinging on varying perceptions (positive, negative, technical etc.) of the term.

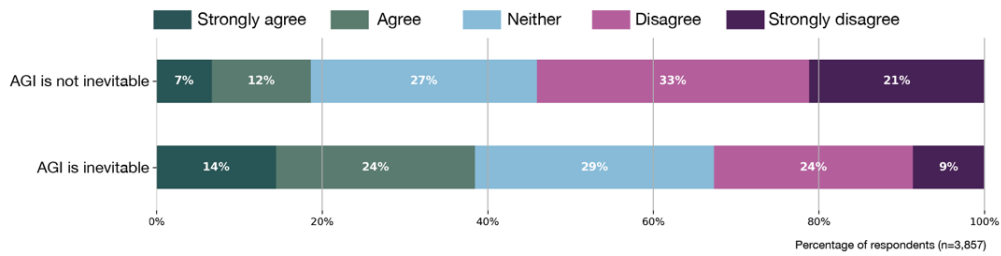

Figure 9. 'To what extent do you agree or disagree: AI should be developed as quickly as possible'

*3.4.5 Sectoral differences*

We were interested in whether the views of researchers differed based on whether they worked in universities, firms or the public sector. Researchers working in firms are, in greater numbers, more likely to agree that members of the public worry too much about the risks of AI (Figure 10). However, across sectors, statistically similar numbers of researchers agree that most people do not understand the potential risks of AI and that the best-placed people to understand the risks of AI are computer scientists.

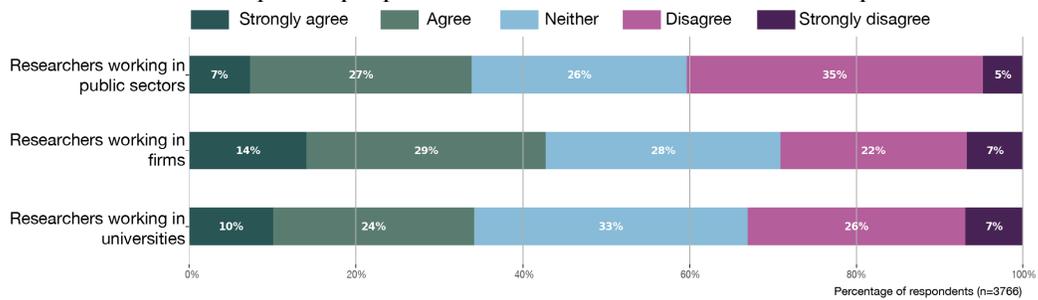

Figure 10. 'Members of the public worry too much about the risks of AI'

Those in firms were more likely to be worried about hype (7% versus 4%, $p = .00002$) and jobs (9% versus 7%). However, we see no significant statistical difference when it comes to considering the impact of AI on different groups in society. There is a significant absence of firm-based respondents expressing worries about the environmental impacts of AI compared to respondents from university/research organizations (0% versus 2%, $p = .00002$; see Table 1 for category definitions).

## 4 DISCUSSION

We should be careful not to read too much into these results. Further qualitative research is needed to explore AI researchers' ideas of risk, opportunity and responsibility. Even the most carefully-designed survey questions will be interpreted differently by each respondent. Respondents will have understood terms in different ways and, where we have made comparisons between researchers' and public views, it should be noted that researchers are likely to have different, more technical understandings of issues. We can assume some self-selection bias in the survey. People who are interested in responding to questions such as ours are unlikely to be perfectly representative of the community of AI researchers. AI researchers' sense of what AI is will also vary. Our respondents here include researchers whose work is highly theoretical and others who focus on uses of particular AI systems.

## 5 CONCLUSIONS

Our survey suggests that there is far more diversity of opinion within AI research than is commonly represented in the public debate. Our survey results support the hypothesis that 'distance lends enchantment' in AI. There is more uncertainty among the researchers who are closest to the technology.

Our survey reveals that, beneath the surface, AI researchers have a range of hopes and fears about the technology that are broader and more complex than the public debate suggests. This diversity, following Callon [6], should be nurtured and made public. Policymakers, journalists and others interested in the public debate on AI should seek more diverse views on AI.

In addition to their views on downstream risk from AI systems, researchers, recognise a range of upstream concerns, such as those to do with the data that feeds AI models, industry control of research agendas and the need for steering of AI research. These are concerns that exist elsewhere in society and as such AI researchers' views seem to be much more 'normal', in the sense used by Narayanan and Kapoor [21], than public representations suggest. Yet current discourse, often dominated by the voices of a small number of powerful interests, risks sidelining insight from researchers, further contributing to public alienation, and acting as a barrier to participation in AI design and decision-making [35]. There is some evidence here that researchers see a division of moral labour, in which scientists are responsible for research and society is responsible for the uses of technology. But there is also evidence here that researchers do not see AI technologies as morally neutral, nor do they see data as merely a raw material. We began this project with interests in the possibilities of democratising AI research. Our survey suggests there is a need for further dialogue, an opportunity to engage AI researchers in a debate that both they and the public find important.

These results suggest a need for further research to explore the connection between researchers' values and priorities and the eventual constitution of AI systems. Our survey results suggest the possibility for new forms of public dialogue on AI's risks and opportunities. Rather than speculating on future potential risks, policymakers and AI researchers should collaborate on understanding and mitigating the range of sociotechnical risks that are already of clear concern. This goes well beyond the dominant sense of AI 'alignment' that imagines both values and AI design narrowly. One of our starting hypotheses was that AI researchers have agency in setting AI trajectories, but other interests, including those of private capital, national security etc., may dominate. We would hope that our sort of survey can become part of the AI research community's understanding of itself, as a tool for reflection and research agenda-setting. However, it is worth noting that more than 75% of our respondents worked in universities. As AI research becomes increasingly dominated by companies [1], whose interests are likely to encourage secrecy, it could become harder to access and hear the voices of AI researchers.

**ACKNOWLEDGEMENTS**

The research reported here was undertaken as part of Public Voices in AI, a satellite project funded by Responsible AI UK and EPSRC (Grant number: EP/Y009800/1). Public Voices in AI was a collaboration between the ESRC Digital Good Network at the University of Sheffield (Grant number: ES/X502352/1), Elgon Social Research Limited, Ada Lovelace Institute, The Alan Turing Institute and University College London.

The project also received support from UCL's Department of Science and Technology Studies, The UCL Centre for Responsible Innovation and the Economic and Social Research Council's Metascience Programme (Grant number: UKRI1087).

**ORCID**
Cian O'Donovan  https://orcid.org/0000-0003-4467-9687
Sarp Gurakan  https://orcid.org/0009-0005-5415-1554
Jack Stilgoe  https://orcid.org/0000-0002-5787-2198